\documentclass{PoS-hep}
\usepackage{amsmath}
\usepackage{bm}
\usepackage{epsfig}
\usepackage{graphics}
\usepackage{upgreek}
\usepackage{amsfonts}
\usepackage{amsbsy}
\usepackage{amscd}
\usepackage{bbm}
\usepackage{eufrak}
\usepackage{cite}
\usepackage{times}
\usepackage{rotating}

\renewenvironment{subequations}{%
\refstepcounter{equation}%
\setcounter{parentequation}{\value{equation}}%
  \setcounter{equation}{0}
  \ignorespaces
}{%
  \setcounter{equation}{\value{parentequation}}%
  \ignorespacesafterend
}
\newcommand{\eeq}{\end{equation}}
\newcommand{\beq}{\begin{equation}}
\newcommand{\ba}{\begin{array}}
\newcommand{\ea}{\end{array}}
\newcommand{\bea}{\begin{eqnarray}}
\newcommand{\eea}{\end{eqnarray}}
\newcommand{\baq}{\begin{eqnarray}}
\newcommand{\eaq}{\end{eqnarray}}

\newcommand{\beqs}{\begin{subequations}}
\newcommand{\eeqs}{\end{subequations}}
\newcommand{\eec}{\end{center}}
\newcommand{\bec}{\begin{center}}
\newcommand{\eem}{\end{matrix}}
\newcommand{\bem}{\begin{matrix}}
\newcommand{\Eref}[1]{Eq.~(\ref{#1})}
\newcommand{\Sref}[1]{Sec.~\ref{#1}}
\newcommand{\Fref}[1]{Fig.~\ref{#1}}
\newcommand{\Tref}[1]{Table~\ref{#1}}
\newcommand{\cref}[1]{Ref.~\cite{#1}}

\newcommand{\etal}{{\it et al.\/}}

\newcommand\eqs[2]{Eqs.~(\ref{#1}) and (\ref{#2})}

\newcommand{\sFref}[2]{Fig.~\ref{#1}-{\ftn\sf ({#2})}}

\newcommand{\ftn}{\footnotesize}

\newcommand{\TeV}{{\mbox{\rm TeV}}}

\newcommand{\GeV}{{\mbox{\rm GeV}}}

\newcommand{\EeV}{{\mbox{\rm EeV}}}
\newcommand{\PeV}{{\mbox{\rm PeV}}}
\def\to{\rightarrow}

\def\lf{\left(}
\def\rg{\right)}
\newcommand\vev[1]{\langle {#1} \rangle}

\newcommand{\Gr}{\ensuremath{\widetilde{G}}}

\newcommand{\Vhi}{\ensuremath{V_{\rm I}}}
\newcommand{\Hhi}{\ensuremath{H_{\rm I}}}

\newcommand{\mP}{\ensuremath{m_{\rm P}}}

\newcommand{\la}{\ensuremath{\lambda_a}}

\def\openone{\leavevmode\hbox{\small1\kern-3.8pt\normalsize1}}
\newcommand{\dV}{\ensuremath{\Delta\widehat V_{\rm I}}}

\newcommand{\Gsn}{\ensuremath{\what{\Gamma}_{\rm \dph}}}

\newcommand{\msn}{\ensuremath{m_{z}}}

\newcommand{\hd}{{\ensuremath{H_d}}}
\newcommand{\hu}{{\ensuremath{H_u}}}

\newcommand{\ks}{\ensuremath{k_\star}}
\newcommand{\Ns}{\ensuremath{{N_\star}}}
\newcommand{\ns}{\ensuremath{n_{\rm s}}}

\newcommand{\as}{\ensuremath{a_{\rm s}}}
\newcommand{\As}{\ensuremath{A_{\rm s}}}

\newcommand{\Ve}{\ensuremath{V}}

\def\al{{\alpha}}
\def\bt{{\beta}}

\def\n{\bar{n}}

\def\th{{\theta}}

\newcommand{\Trh}{\ensuremath{T_{\rm rh}}}
\newcommand{\sg}{\ensuremath{z}}

\newcommand{\Ld}{\ensuremath{\Lambda}}

\newcommand{\sgx}{\ensuremath{z_\star}}
\newcommand{\sgf}{\ensuremath{z_{\rm f}}}

\newcommand{\what}{\ensuremath{\widehat}}
\newcommand{\wtilde}{\ensuremath{\widetilde}}
\newcommand{\mss}{\ensuremath{\widetilde m}}

\newcommand{\se}{\ensuremath{\widehat z}}
\newcommand{\sex}{\ensuremath{\widehat{z}_\star}}
\newcommand{\sef}{\ensuremath{\widehat{z}_{\rm f}}}

\newcommand{\mgr}{\ensuremath{m_{3/2}}}
\newcommand{\mz}{\ensuremath{ m_{z}}}
\newcommand{\mth}{\ensuremath{m_{\theta}}}

\def\Ka{K\"{a}hler potential}

\def\Km{K\"{a}hler manifold}
\def\Kaa{K\"{a}hler~}

\newcommand{\plk}{{\it Planck}}

\newcommand{\Ggr}{\ensuremath{{\Gamma}_{3/2}}}
\renewcommand{\Gsn}{\ensuremath{{\Gamma}_{\dzv}}}
\newcommand{\Gth}{\ensuremath{{\Gamma}_{\theta}}}
\newcommand{\Gh}{\ensuremath{{\Gamma}_{\tilde{h}}}}

\newcommand{\Vs}{\ensuremath{V_{\rm I\star}}}
\newcommand{\Hs}{\ensuremath{H_{\rm I\star}}}

\newcommand{\phc}{\ensuremath{\Phi}}

\newcommand{\wo}{\ensuremath{W_0}}
\newcommand{\wopm}{\ensuremath{W_0^{\pm}}}
\newcommand{\wop}{\ensuremath{W_0^{+}}}
\newcommand{\wom}{\ensuremath{W_0^{-}}}
\newcommand{\wcc}{\ensuremath{W_{\Lambda}}}
\newcommand{\gk}{\ensuremath{g_{K}}}
\newcommand{\vcc}{\ensuremath{V_{\Lambda}}}

\newcommand{\cm}{\ensuremath{C_{\Lambda}}}
\newcommand{\clup}{\ensuremath{C_{u\boldsymbol+}^-}}
\newcommand{\clum}{\ensuremath{C_{u\boldsymbol-}^-}}
\newcommand{\clw}{\ensuremath{C_{\omega}^-}}

\newcommand{\vrm}{\ensuremath{{\rm v}}}
\newcommand{\vp}{\ensuremath{v_{\boldsymbol+}}}
\newcommand{\vm}{\ensuremath{v_{\boldsymbol-}}}

\newcommand{\up}{\ensuremath{u_{\boldsymbol+}}}
\newcommand{\um}{\ensuremath{u_{\boldsymbol-}}}

\newcommand{\zv}{\ensuremath{Z_{\rm v}}}
\newcommand{\tv}{\ensuremath{T_{\rm v}}}

\newcommand{\nm}{\ensuremath{n_{\boldsymbol -}}}
\newcommand{\nnp}{\ensuremath{n_+}}
\newcommand{\nnm}{\ensuremath{n_-}}
\newcommand{\nnpm}{\ensuremath{n_\pm}}
\newcommand{\nnmp}{\ensuremath{n_\mp}}

\newcommand{\cfm}{\ensuremath{C_{f}^-}}

\newcommand{\ctm}{\ensuremath{C_{T}^-}}

\newcommand{\Om}{\ensuremath{\Omega}}
\newcommand{\om}{\ensuremath{\omega}}
\newcommand{\omz}{\ensuremath{\Omega_{,Z}}}
\newcommand{\omzz}{\ensuremath{\Omega_{,ZZ^*}}}

\newcommand{\Ctp}{\ensuremath{C_{\om}^+}}
\newcommand{\Ctm}{\ensuremath{C_{\om}^-}}

\def\bz{{Z^*}}
\def\bzz{{\bar z}}

\newcommand{\sgo}{\ensuremath{z_0}}
\newcommand{\ko}{\ensuremath{k_0}}
\renewcommand{\sgx}{\ensuremath{z_\star}}
\renewcommand{\sgf}{\ensuremath{z_{\rm f}}}

\newcommand{\dz}{\ensuremath{\delta z}}
\newcommand{\dzx}{\ensuremath{\delta z_{\star}}}
\newcommand{\dzf}{\ensuremath{\delta z_{\rm f}}}
\newcommand{\dk}{\ensuremath{\delta k}}
\newcommand{\dzv}{\ensuremath{{\updelta} z}}
\newcommand{\dzvh}{\ensuremath{\what{\updelta z}}}

\newcommand{\vo}{\ensuremath{v_{0}}}
\newcommand{\va}{\ensuremath{v_{1}}}
\newcommand{\vb}{\ensuremath{v_{2}}}
\newcommand{\vc}{\ensuremath{v_{3}}}
\newcommand{\jo}{\ensuremath{J_{0}}}

\newcommand{\ep}{\ensuremath{\epsilon}}
\newcommand{\pn}{\ensuremath{p_{N}}}
\newcommand{\fns}{\ensuremath{f_{N\star}}}
\newcommand{\fnf}{\ensuremath{f_{N{\rm f}}}}
\newcommand{\fn}{\ensuremath{f_{N}}}

\newcommand{\bdhh}{{\ensuremath{\normalsize I{\kern-2.9pt H}}}}

\title{\boldmath \bfseries High-Scale SUSY from Sgoldstino
Inflation}

\ShortTitle{High-Scale SUSY from Sgoldstino Inflation}

\author{\speaker{C. Pallis}\\
School of Civil Engineering, \\ Faculty of Engineering, \\
Aristotle University of Thessaloniki, \\ GR-541 24 Thessaloniki, GREECE \\
E-mail: \email{kpallis@auth.gr}}

\abstract{We review a number of unimodular no-scale supergravity
models with F-term SUSY breaking which support technically natural
de Sitter vacua. A variant of these models develops a stage of
inflection-point inflation which can be realized for subplanckian
field values consistently with the observational data. For central
value of the spectral index $\ns$, the necessary tuning is of the
order of $10^{-6}$, the tensor-to-scalar ratio is tiny whereas the
running of $\ns$ is around $-3\cdot10^{-3}$. Our proposal is
compatible with high-scale SUSY and the results of LHC on the
Higgs boson mass.
\\ \\{\sl\bfseries Published in}~~{PoS  CORFU {\bf 2023}, 055 (2024)}.

}

\FullConference{Corfu Summer Institute 2023 "School and Workshops on Elementary
Particle Physics and Gravity" (CORFU2023) 23 April - 6 May and 27 August - 1 October, 2023\\
Corfu, Greece}

\begin{document}


\section{Outline}

In this talk we first -- see \Sref{sec1} -- review a set of
(generalized) \emph{no-scale models} ({\sf\ftn nSMs}) \cite{ns89,
de} within \emph{Supergravity} ({\sf\ftn SUGRA}) which assure
spontaneous \emph{Supersymmetry} ({\sf\ftn SUSY}) breaking at a
technically natural \emph{de Sitter} ({\sf\ftn dS}) vacuum. As a
result, the problem of \emph{Dark Energy} ({\sf\ftn DE}) can be
explained by fine tuning just one superpotential coupling. We
then, in \Sref{sec2}, concentrate on a specific model which offers
a coexistence of the aforementioned dS vacuum with an inflection
point of the potential developed at larger field values and, in
\Sref{sec3}, we investigate the communication of SUSY breaking in
the observable sector of \emph{minimal SUSY standard model}
({\sf\ftn MSSM}). Finally, in \Sref{sec4} we show how we can
obtain \emph{Inflection-Point Inflation} ({\sf\ftn IPI})
\cite{dei} in our set-up and in \Sref{res} we delineate regions of
parameters allowed by the observations \cite{plin,plcp,bk15}.
\Sref{con} summarizes our conclusions and discusses some open
issues.

Unless otherwise stated, we use units where the reduced Planck
scale $\mP=2.4\cdot 10^{18}~\GeV$ is taken to be unity, a
subscript of type $,\chi$ denotes derivation \emph{with respect
to} ({\small\sf w.r.t.}) the field $\chi$ and charge conjugation
is denoted by a star.


\section{From Minkowski to {\bf d}S Vacua In No-Scale SUGRA}\label{sec1}

We here provide a short introduction on SUSY breaking within SUGRA
in \Sref{sec1a} and then, we show how this mechanism is
systematized within no-scale SUGRA obtaining Minkowski (see
\Sref{sec1b}) or dS vacua -- see \Sref{sec1c}. Examples of such
nSMs are given in \Sref{sec1d}.

\subsection{SUSY breaking within SUGRA}\label{sec1a}

Within global SUSY, the scalar potential of a {gauge-singlet}
superfield $Z$, $V_{\rm SUSY}$, is positive semi-definite, since
\beq V_{\rm SUSY}=|F_Z|^2~~\mbox{with}~~F_Z = \partial_Z W. \eeq
Here $W=W(Z)$ is an holomorphic function named {superpotential}.
Spontaneous SUSY breaking occurs when $\vev{F_Z}\neq0$ which
results to $\vev{V_{\rm SUSY}}^{1/4}>0$. The non discovery of SUSY
in LHC dictates $\vev{F_Z}^{1/2}>1~\TeV$. On the other hand,
$\vev{F_Z}^{1/2}$ may be identified with the cosmological constant
$\Lambda_{\rm CC}\simeq2.3~{\rm meV}$. Therefore, we obtain an
inconsistency. The spontaneous SUSY breaking is accompanied with
the presence of a massless fermion named {goldstino} and for this
reason its SUSY partner, $Z$, is called {sgoldstino}.

Within local SUSY -- i.e. SUGRA -- the F-term scalar potential is
given by
\beq  V_{\rm SUGRA}=e^{G}\lf G^{ZZ^*} G_Z
G_{Z^*}-3\rg~~\mbox{where}~~ G := K + \ln |W|^2\label{vsugra}\eeq
is the K\"{a}hler-invariant function and $K=K(Z,Z^*)$ the \Ka.
Also $G_{ZZ^*} =K_{ZZ^*}=\partial_Z\partial_{Z^*} K$ is the {\Kaa
metric} and $K^{ZZ^*}=K_{ZZ^*}^{-1}$. In this context, SUSY is
broken again when $\vev{F^Z}\neq0$ where $F^Z =
e^{G/2}K^{ZZ^*}G_{Z^*}$ which may occur with $\vev{V_{\rm
SUGRA}}\simeq0$. This mechanism is accompanied with the absorption
of the goldstino by the {gravitino}, $\Gr$, which acquires mass
according to ``super-Higgs'' mechanism
\beq \mgr=\vev{e^{G/2}}=\vev{e^{K/2}W}=\vev{G_{Z\bz}F^Z\bar
F^\bz-V_{\rm SUGRA}}^{1/2}/\sqrt{3}.\label{mgr}\eeq

Based on this formalism we can obtain several models of SUSY
breaking \cite{polonyi, kallosh, susyr, hall}. One of the key
ingredients for the successful implementation of this scenario is
the determination of a naturally realistic vacuum for $V_{\rm
SUGRA}$ which may be Minkowski for $\vev{V_{\rm SUGRA}}=0$ or de
Sitter for $\vev{V_{\rm SUGRA}}>0$.
\subsection{Minkowski Vacua In no-scale SUGRA}\label{sec1b}

Within no-scale SUGRA \cite{ns89}, SUSY is broken along an F-flat
direction which naturally yields $V_{\rm SUGRA}=0$. To construct
systematically nSMs, we use as input $K$ and determine $W$ so as
$V_{\rm SUGRA}=0$. I.e., we solve the differential equation
\beq 0= e^K\lf \gk^{-1}\left |\partial_Z W+W
K_Z\right|^2-3|W|^2\rg,~~\mbox{where}~~\gk^{-1}=K_{ZZ^*}^{-1}=K^{ZZ^*}\label{Vs0}\eeq
assuming that the direction $Z=Z^*$ is stable -- this assumption
can be verified a posteriori. Indeed, solving \Eref{Vs0} w.r.t
$W=\wo(Z)$ we find
\beq \frac{d\wo}{dZ\wo}=\pm\sqrt{3\gk}-K_Z~~\Rightarrow~~
\wopm=m\exp\lf \pm\int dZ\sqrt{3\gk}-\int
dZK_Z\rg.\label{wopm}\eeq
with $'=d/dZ$. E.g., if we select the \Ka:

\begin{itemize}

\item  $K=-3\ln(T+T^*)$ we obtain the well-known results
\cite{nsreview} $W_0^-=m$ but also $W_0^+=8mT^3$ ;

\item $K=|Z|^2$ we obtain  $\wopm=me^{\pm\sqrt{3}Z-Z^2/2}$.
Therefore we can obtain a nSM even with flat geometry. This is a
totally novel result \cite{de}.

\end{itemize}

\subsection{{\bf d}S Vacua In no-Scale SUGRA}\label{sec1c}

The models can be extended to support dS vacua. In this case
$\vev{V_{\rm SUGRA}}$ may account for DE without requiring any
external mechanism for vacuum uplifting \cite{kallosh}. To
accomplish this extra achievement we consider the following linear
combination of $\wopm$
\beq \wcc=\wop-\cm\wom\label{wcc}\eeq
and we obtain the SUGRA potential
\beq\vcc= e^K\lf \gk^{-1}\lf\wcc'+\wcc K_Z\rg^2 -3\wcc^2\rg
=12e^K\cm\wom\wop=12m^2\cm.   \label{vcc}\eeq
Finely tuning $\cm$ to a value $\cm\simeq10^{-108}$ for
$m\sim10^{-6}$, e.g, we may identify $\vcc$ with the present DE
energy density, i.e.,
\beq \label{omde} \vcc=\Omega_\Lambda\rho_{\rm
c}=7.3\cdot10^{-121}\mP^4,\eeq
where the density parameter of DE $\Omega_\Lambda$ and the current
critical energy density of the universe $\rho_{\rm c}$ are given
in \cref{plcp}.
%

\subsection{Realistic nSMs}\label{sec1d}

Although quite appealing, the nSMs above develop a completely flat
$V_{\rm SUGRA}$, I.e.
\beq V_{\rm SUGRA}=V'_{\rm SUGRA}=V''_{\rm SUGRA}=0. \eeq
Therefore $\mgr$ and the soft SUSY-breaking terms remain
undetermined. Moreover, a massless mode arises in the spectrum. To
cure these drawbacks, we may include in $K$ a {stabilization
(higher order) term}
\beq-k^2\zv^4~~\mbox{with}~~\zv=Z+Z^*-\sqrt{2}\vrm\eeq
which selects the vacuum $(\vev{z},\vev{\bzz})=(\vrm,0)$ from the
flat direction and provides the real component of sgoldstino with
mass. The presence of $k$-dependent term is natural according to
't Hooft argument \cite{symm} since for $k\to 0$ the symmetry
becomes exact. The selection of this higher order term is
arbitrary, though.

\newcommand{\art}{{\rm atn}}
\newcommand{\arth}{{\rm atnh}}

\begin{sidewaystable}[h]
\renewcommand{\arraystretch}{1.4}
\bec
\begin{tabular}{|c|c|c|c|c|}\hline
n{\sc SM} &$K$&$\wopm/m$&$W_\Ld/m$&{\sc Enhanced Symmetry}\\
&&&&{\sc of the \Kaa\ Manifold} \\\hline\hline
1&$-N\ln\lf T+T^*+k^2\tv^4/N\rg,$&$(2T)^{\nnmp},$ where &$(2T)^{\nnp}\ctm,$ where& $SU(1,1)/U(1)$\\
&$\tv=T+T^*-\sqrt{2}\vrm,
N>0$&$\nnpm=(N\pm\sqrt{3N})/2$&$\ctm=1-\cm
(2T)^{-\sqrt{3N}}$&{i.e., Hyperbolic Geometry with}\\\cline{1-4}
2&${\boldmath-}N\ln\lf 1{\boldmath
-}|Z|^2/N+k^2\zv^4/N\rg,$&$\vm^{-N/2}\um^{\pm1},$ where
&$\vm\um\clum,$ where&{$\bullet$ Half-plane
coordinates for nSM1~~~~~}\\
&$\zv=Z+Z^*-\sqrt{2}\vrm,
N>0$&\multicolumn{1}{|c|}{$\vm=1{\boldmath -}Z^2/N$}& $\clum=1-\cm
\um^{-2}$&{$\bullet$ Poincar\'e-disc coordinates for
nSM2}\\\cline{4-4}
&&\multicolumn{1}{|c|}{$\um=e^{\sqrt{3N}{\boldmath
\arth}(Z/N)}$}&$\arth:={\rm arctanh}$&{}\\\hline
3&${\boldmath+}N\ln\lf
1{\boldmath +}|Z|^2/N-k^2\zv^4/N\rg,$&$\vp^{-N/2}\up^{\pm1},$ where&$\vp\up\clup,$ where&$SU(2)/U(1)$\\
&$\zv=Z+Z^*-\sqrt{2}\vrm,
N>0$&\multicolumn{1}{|c|}{$\vp=1{\boldmath +}Z^2/N$}&$\clup=1-\cm
\up^{-2}$&{i.e., Compact Geometry}\\\cline{4-4}
&&\multicolumn{1}{|c|}{$\up= e^{\sqrt{3N}{\boldmath
\art}(Z/N)}$}&$\art=\arctan$&{}\\\hline
4&$|Z|^2-k^2\zv^4,$&$wf^{\pm1},$ where &$wf\cfm,$ where&$U(1)$\\
&$\zv=Z+Z^*-\sqrt{2}\vrm$&\multicolumn{1}{|c|}{$w=e^{-Z^2/2}$ and
$f=e^{\sqrt{3}Z}$}&$\cfm=1-\cm f^{-2}$&{i.e., Flat
Geometry}\\\hline
\end{tabular}\eec
\caption{\sl\small Uni-Modular no-Scale Models (nSMs) with {dS}
vacua.}\label{tab1}
\renewcommand{\arraystretch}{1.}
\end{sidewaystable}
\clearpage

Applying the procedure above several nSMs with {\rm dS} vacua can
be established varying the \Kaa\ geometry. In \Tref{tab1} we
arrange a catalogue of such models based exclusively on one
modulus. These models are introduced in \cref{de} where
multi-moduli models are also exposed -- see also \cref{ns89}. For
each nSM we can see there the adopted \Ka, the solutions of
\Eref{wopm} and the resulting $\wcc$ from \Eref{wcc}. The enhanced
symmetry (for $k\to0$) of the \Km\ is also shown in the rightmost
column of \Tref{tab1}. For $k=0$ the \Km\ of nSM1 enjoys the
$SU(1,1)/U(1)$ hyperbolic symmetry parameterized by the half-plane
coordinates $T$ and $T^*$. As a result, the expression in
\Eref{wopm} is a polynomial of $2T$. For the well-known nSM with
$N=3$ \cite{nsreview}, we obtain $\nnp=3$ and $\nm=0$ and so, we
have the ingredients
\beq
K=-3\ln(T+T^*)\>\>\mbox{and}\>\>W_\Ld=8mT^3\ctm\>\>\mbox{where}\>\>\ctm=1-\cm(2T)^{-3}.
\eeq
The same enhanced symmetry parameterized in the Poincar\'e-disc
coordinates $Z$ and $Z^*$ is valid for nSM2. That parametrization
allows us to pass from the non-compact to the compact geometry of
nSM3 by changing the signs in the relevant $K$'s. As a consequence
we obtain a remarkable correspondence between nSM2 and nSM3 as
regards the relevant expressions of $W_0^{\pm}$. Namely $\vm$ and
${\rm arctanh}$ in nSM2 are replaced by $\vp$ and $\arctan$ in
nSM3. Finally, we consider nSM4 where a flat \Ka\ is adopted
resulting to exponential $W_0^{\pm}$.

To check the stability of the vacuum of the nSMs in \Tref{tab1},
we derive the mass spectrum at the vacuum. The results are
presented in \Tref{tab2}. We remark that we need $k>0$ and $N>3$
when the (enhanced) \Kaa\ geometry is hyperbolic. In a such case
we also obtain real values for the mass $m_{\bzz}$ of the
imaginary component of $Z$. Note that we here  decompose $Z$ as
$Z=(z+i\bzz)/\sqrt{2}$.

\renewcommand{\arraystretch}{1.3}
\begin{table}[t] \bec\begin{tabular}{|c|c|c|c|c|}\hline
&\multicolumn{2}{c|}{\sc Mass Of Sgoldstino
Components}&$\mgr$&{\sc Restriction}\\\cline{2-3}
n{\sc SM}&{\sc Real}&{\sc Imaginary}&&\\\hline\hline
1&$24k\vrm^{3/2}\mgr$&$2(1-3/N)^{1/2}\mgr$&$m(2\vrm^2)^{\sqrt{3N}/4}$&{$N>3$}\\\hline
2&$12k\vev{\vm}^{3/2}\mgr$&$2(1-3/N)^{1/2}\mgr$&$m\vev{\um}$&{$N>3$}\\\hline
3&$12k\vev{\vp}^{3/2}\mgr$&$2(1+3/N)^{1/2}\mgr$&$m\vev{\up}$&-\\\hline
4&$12k\mgr$&$2\mgr$&$me^{\sqrt{3/2}\vrm}$&-\\\hline
\end{tabular}\eec
\caption{\sl\small  Particle mass spectrum at the vacuum for the
nSMs presented in Table 1.}\label{tab2}
\end{table}
\renewcommand{\arraystretch}{1.}

\section{DE and Inflection Point From no-Scale SUGRA}\label{sec2}

We aspire to identify the radial component of the sgoldstino $Z$
with the inflaton -- for similar attempts see \cref{ant1}. We
accomplish this aim by localizing an inflection point of the
potential $V_{\rm SUGRA}(Z)$. Close to it we may obtain a stage of
IPI according to formalism discussed in \cref{ipisusy, drees}. To
achieve it for $z>\vrm$ we adopt nSM1 in \Tref{tab1} with
$T=1/2-Z/2$. I.e., we set
\beq
K=-N\ln\Omega\>\>\>\mbox{with}\>\>\>N>0,\>\>\Omega=1-(Z+Z^*)/2+k^2\zv^4\>\>\>\mbox{and}\>\>\>\zv=Z+Z^*-2\vrm.\eeq
For $k=0$, $K$ enjoys a symmetry related to a subset of $U(1,1)$
without to define specific \Kaa\ manifold \cite{epole}. Repeating
the procedure in \Sref{sec1}, we find that $K$ may be associated
with
\beq
\label{wccn}W_{\Lambda}=m\om^{\nnp}\clw\>\>\>\mbox{with}\>\>\>\nnp=(N+\sqrt{3N})/2,\eeq
\beq\om=\Omega(Z=Z^*,k=0)=1-Z\>\>\>\>
\mbox{and}\>\>\clw=1-\cm\om^{-\sqrt{3N}}\label{om}\eeq
where $m$ is an arbitrary mass scale which is constrained to
values close to $10^{-7}$ from the normalization of $\As$ -- see
below.

\begin{figure}[t]\vspace*{-6.1cm}
\begin{minipage}{85mm}
\includegraphics[width=8.7cm,angle=-0]{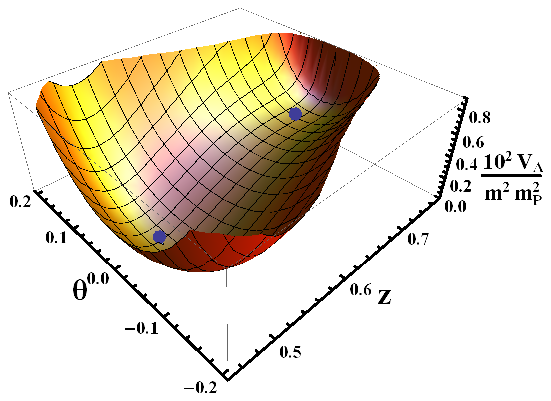}
\end{minipage}
\begin{minipage}{85mm} \begin{center}
\vspace*{3.1in}\begin{tabular}{|c|c|c|}\hline
$m/\mP$&$\cm/10^{-108}$&$k/0.1$\\\hline
$5.6\cdot10^{-7}$&$2.5$&$4.0167291$\\\hline
$\vrm/\mP$&$N$&$(\nnp,\nnm)$\\\hline
$0.5$&$12$&$(9,3)$\\\hline \end{tabular}
\end{center}
\end{minipage}
\vspace*{.6in}\caption{\sl The (dimensionless) SUGRA potential
$10^2\vcc/m^2\mP^2$ in Eq.~(3.5) as a function of $z$ and $\th$
defined above Eq. (3.6). The location of the dS vacuum at
$(\vev{z},\th)=(0.5,0)$ and the
inflection point at $(\sgo,\th)\simeq(0.71,0)$ is also depicted by two thick 
points. The input parameters are listed in the Table.}
\label{fig1}\end{figure}

The exponents $\nnp$ in \Eref{wccn} may, in principle, acquire any
real value, if we consider $W_{\Lambda}$ as an effective
superpotential. When $N/3>1$ is a perfect square, integer $\nnpm$
values may arise too. E.g.
\beq \mbox{For $N=12, 27$ and $48$ we obtain $(\nnm,\nnp)=(3,9),
(9,18)$ and $(18,30)$ respectively.} \eeq
The resulting SUGRA potential is \cite{dei}
\beq \vcc(Z)=m^2\Om^{-N}\om^{2\nnp}
\lf|U/2\om|^2-3|\Ctm|^2\rg,\label{vccn}\eeq
where we define the auxiliary quantity
\beq
U=\frac{\sqrt{2N}}{J\Om}\Bigg(\lf\sqrt{3}\Ctp+\sqrt{N}\Ctm\rg\Om+2\sqrt{N}\Ctm\omz\om\Bigg).
\eeq
The canonically normalized components of the complex scalar field
$Z=ze^{i\theta}$ are
\beq  \label{Jg} \frac{d\widehat
z}{dz}=\sqrt{2K_{ZZ^*}}=J~~\mbox{and}~~ \what{\theta}=
J\sg\theta,\eeq where $J$ can  be expressed in terms of $\Om$ as
follows
\beq
J=\sqrt{2N}\lf\frac{\omz^2}{\Om^2}-\frac{\omzz}{\Om}\rg^{1/2}\>\>
\mbox{with}~~  \omz=
-1/2+4k^2\zv^3~~\mbox{and}~~\omzz=12k^2\zv^2.\eeq

If we plot $10^2\vcc/m^2\mP^2$ as a function of $z$ and $\theta$
for the inputs in shown the Table of \Fref{fig1} we see that
$\vcc$ develops two critical points: The SUSY-breaking dS vacuum
which is
\beq
\label{vevn}\vev{z}=\vrm~~\mbox{and}~~\vev{\theta}=0~~\mbox{with}~~\vev{\vcc}=12\cm
m^2 \eeq
and an inflection point for $z=\sgo\simeq0.71>\vrm$ which lets
open the possibility of an inflationary stage. The vacuum in
\Eref{vevn} is stable against fluctuations of the various
excitations for $N>3$ which assures $m_{\theta}^2>0$ in accordance
with our findings in \Tref{tab2}. Indeed, we find
\beq \label{msa} m_{z}\simeq48\mgr
kN^{-1/2}\vev{\om}^{3/2}\>\>\>\mbox{and}\>\>\>
m_{\theta}\simeq2\mgr\lf1-(3/N)\rg^{1/2}\>\>\>\mbox{with}\>\>\>\mgr=m\vev{\om}^{\sqrt{3N}/2}.\eeq
Numerical values for the masses above are given in \Tref{tab3} and
all of these are of the order of $100~\EeV$.

\begin{table}[t]\bec\begin{tabular}{|c||c|c|c|c|}\hline
$m/\EeV$&$\msn/\EeV$&$m_{\theta}/\EeV$&$\mgr/\EeV$\\\hline
$1344$&$319$&$281$&$162$\\\hline
\end{tabular}\eec
\caption{\sl\small Particle mass spectrum in \EeV
($1~\EeV=10^9~\GeV$) for the inputs in
Fig.~1.}\end{table}\label{tab3}
\renewcommand{\arraystretch}{1.}

\section{SUSY Breaking and High-Scale MSSM}\label{sec3}

The SUSY breaking occurred at the vacuum in \Eref{vcc} can be
transmitted to the visible world if we specify a reference low
energy model. We here adopt MSSM and the total superpotential,
$W_{\rm \Ld O}$ and \Ka\ $K_{\rm  \Ld O}$,  of the theory take the
form \cite{soft}
\beq \label{Who} W_{\rm \Ld O}=W_\Ld(Z) + W_{\rm
MSSM}\lf\phc_\al\rg~~\mbox{and}~~K_{1\rm\Ld
O}=K(Z)+\sum_\al|\phc_\al|^2~~\mbox{or}~~K_{2\rm\Ld O}=K(Z)+N_{\rm
O}\ln\left(1+\sum_\al\frac{|\phc_\al|^2}{N_{\rm O}}\right)\eeq
where $N_{\rm O}$ may remain unspecified and $W_{\rm MSSM}$ has
the well-known form written in short as
\beq W_{\rm MSSM}=h_{\al\bt\gamma} \phc_\al\phc_\bt\phc_\gamma/6
+\mu\hu\hd ~~\mbox{with}~~\phc_\al= {Q}, {L}, {d}^c, {u}^c, {e}^c,
\hd~~\mbox{and}~~\hu\label{wo}\eeq
the various chiral and Higgs superfields -- we suppress the
generation indices for simplicity. We also denote the three
non-vanishing Yukawa coupling constants as
$$h_{\al\bt\gamma}=h_D, h_U ~~\mbox{and}~~h_E~\mbox{for}~~
(\al,\bt,\gamma)=(Q,\hd,d^c),
(Q,\hu,u^c)~~\mbox{and}~~(L,\hd,e^c)$$
respectively. Also, working in the regime of high-scale SUSY
\cite{strumia} $\mu$ acquires values close to $\mgr$ and we handle
it as a free parameter.

Adapting the general formulae of \cref{soft}, we find universal
(i.e., $\wtilde m_\al=\mss$ and $A_{\al\bt\gamma}=A$) soft
SUSY-breaking terms in the effective low energy potential which
can be written as
\beq V_{\rm SSB}=\wtilde m^2 |\phc_\al|^2+\lf\frac16 A\what
h_{\al\bt\gamma} \phc_\al\phc_\bt\phc_\gamma+ B\what\mu
H_uH_d+{\rm h.c.}\rg ~~\mbox{with}~~(\what
h_{\al\bt\gamma},\what\mu)=\vev{\om}^{-N/2}(
h_{\al\bt\gamma},\mu),\label{vmssm} \eeq
the normalized (hatted) parameters. Also the soft SUSY-breaking
parameters are found to be
\beq\label{mAB} \wtilde m=\mgr,~~|A|=\sqrt{3N}\mgr~~\mbox{and}~~
|B|=(1+\sqrt{3N})\mgr.\eeq

As regards the gauginos of MSSM we expect that we can obtain
similar values selecting the gauge-kinetic function as\cite{soft}
\beq f_{\rm a}= \la Z\, \label{gkf} \eeq
where $\la$  is a free parameter absorbed by a redefinition of the
relevant spinors and ${\rm a}=1,2,3$ runs over the factors of the
gauge group of MSSM, $U(1)_Y$, $SU(2)_{\rm L}$ and $SU(3)_{\rm c}$
respectively. In a such case, we find \cite{soft}
\beq |M_{\rm a}|= \sqrt{3/N}\vev{\om/z}\mgr,\label{Ma} \eeq
which is obviously of the order of $\mgr$.

\begin{table}[t]\bec \begin{tabular}{|c|c|c|c|c|c|}\hline
$m/\EeV$&\multicolumn{1}{|c||}{$\what\mu/\EeV$}&$\mss/\EeV$&$|A|/\EeV$&$|B|/\EeV$&$|M_{\rm
a}|/\EeV$\\\hline
$1344$&\multicolumn{1}{|c||}{$81$}&$162$&$1024$&$1200$&$81.1$\\\hline
\end{tabular}\eec
\caption{\sl\small Soft SUSY-breaking parameters in \EeV
($1~\EeV=10^9~\GeV$) for the inputs in Fig.~1.}\label{tab4}
\end{table}

Representative values for the soft SUSY parameters are displayed
in \Tref{tab4} for $\what\mu=\mgr/2$. We see that $|A|>\mgr$ and
$|B|>\mgr$  due to large $N$ adopted there. However, these
parameters have very suppressed impact on the SUSY mass spectra.
Scenarios with large $\mss$, although not directly accessible at
the LHC, can be probed via the measured value of the Higgs boson
mass. Within high-scale SUSY, updated analysis requires
\cite{strumia}
\beq 3\lesssim\mss/\EeV\lesssim300,\label{highb} \eeq
for degenerate sparticle spectrum, low $\tan\beta$ values and
minimal stop mixing. From the values in \Tref{tab4} we conclude
that our setting is comfortably compatible with the requirement
above.

\section{Inflection-Point Inflation (IPI)}\label{sec4}

The inflection point developed at large values of $\Vhi$ opens up
the possibility of the establishment of IPI \cite{dei} -- cf.
\cref{ipisusy}. We below show how we can systematize the
specification of this inflection point in \Sref{sec4a}, present
the inflationary dynamics and outputs consistently with the
reheating stage occurring after IPI.

\subsection{Localization of the Inflection-Point}\label{sec4a}

\begin{figure}[!t]\vspace*{-.12in}
\hspace*{-.12in}
\begin{minipage}{8in}
\epsfig{file=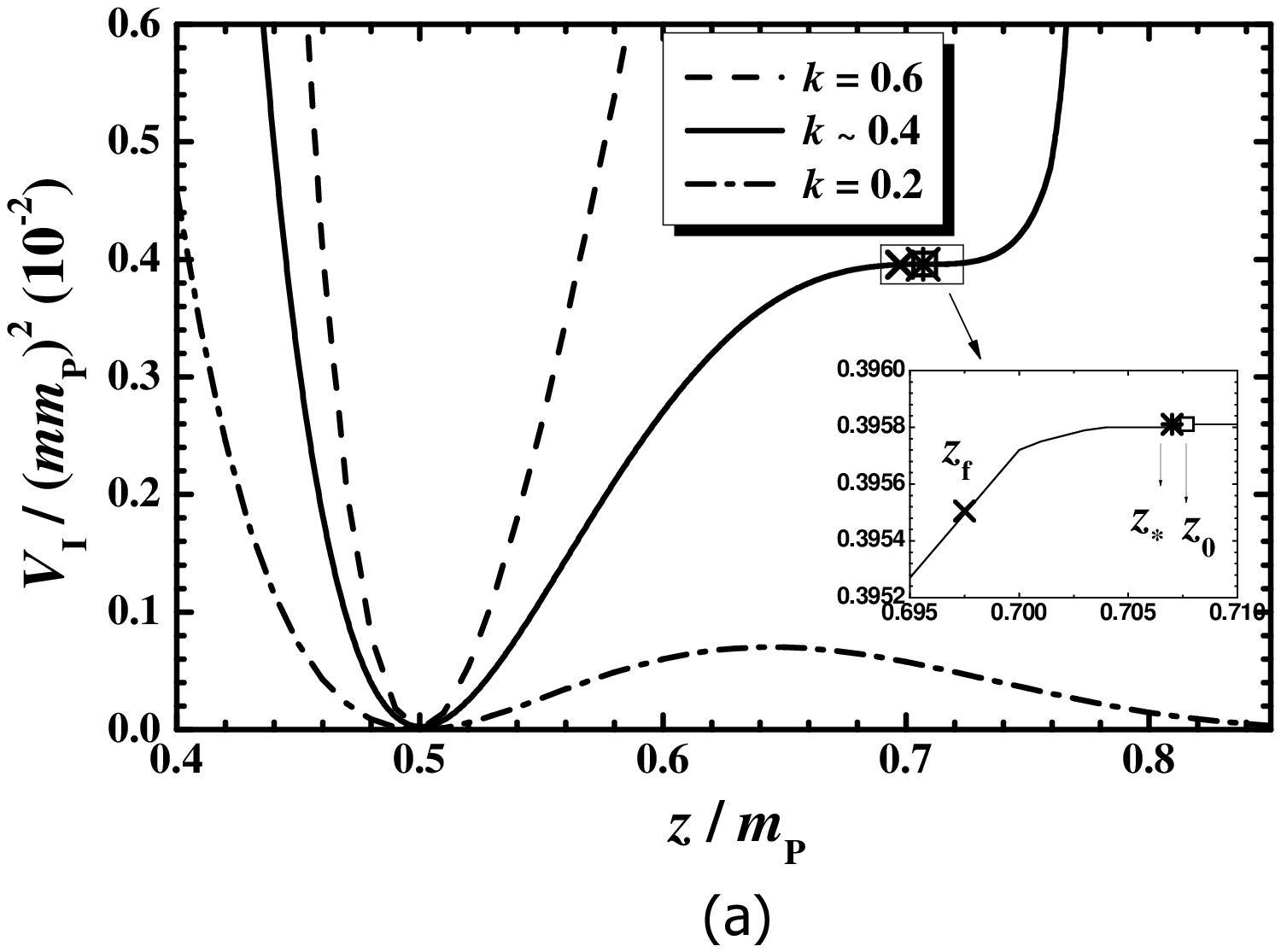,height=3.3in,angle=-90} \hspace*{-.2cm}
\epsfig{file=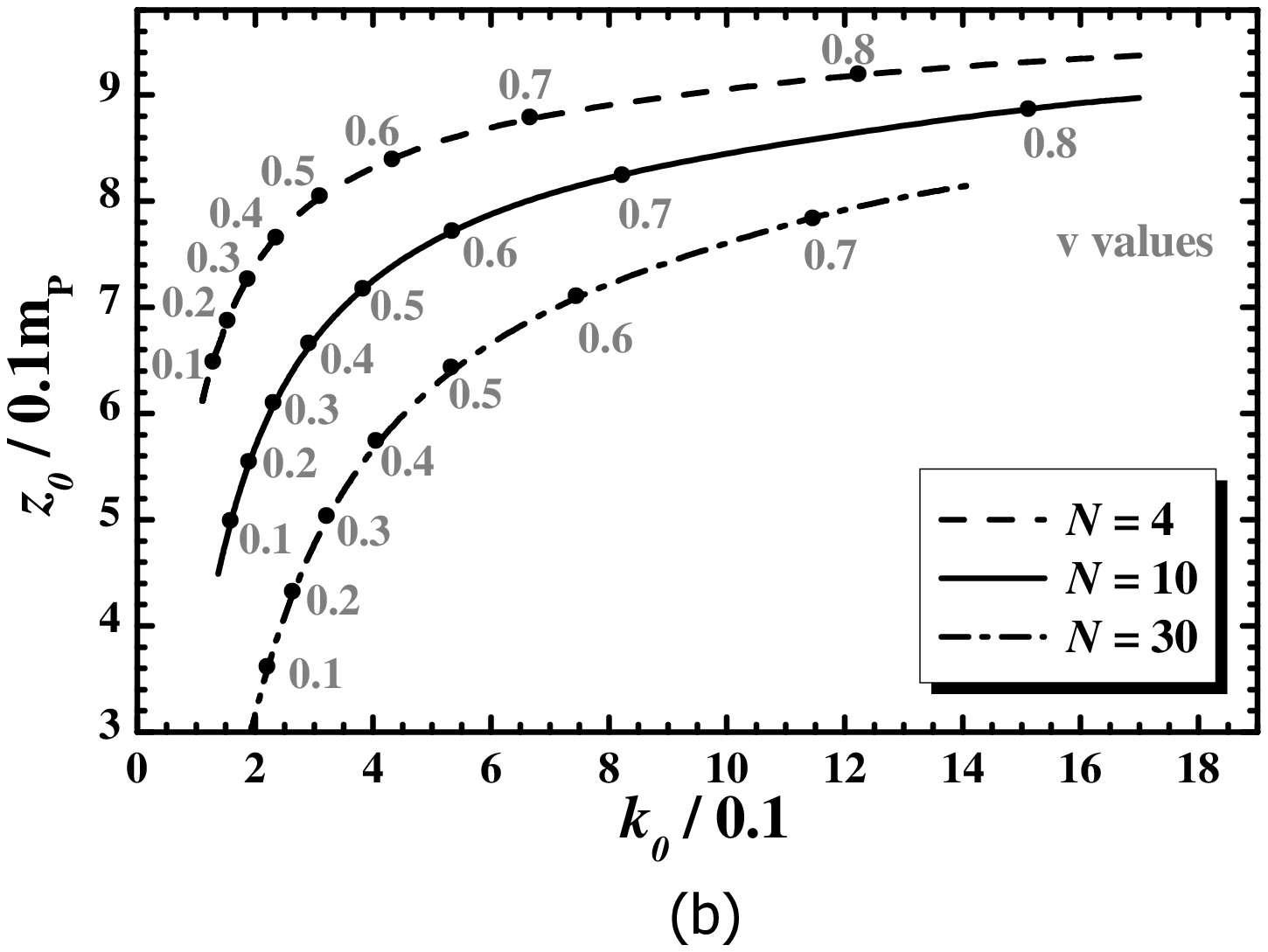,height=3.3in,angle=-90} \hfill
\end{minipage}
\hfill \caption{\sl\small ({\sf\ftn a}) Dimensionless inflationary
potential $\Vhi/m^2\mP^2$ as a function of $\sg$ for
$k=0.40167291$ (black line) or $k=0.2$ (dot-dashed line) or
$k=0.6$ (dashed line) and the remaining inputs in Fig. 1. The
values of $\sgx$, $\sgf$ and $\sgo$ (for the first case) are also
indicated. ({\sf\ftn b}) Location of the inflection point in the
$\ko-\sgo$ plane for various $N$'s indicated in the plot. Shown is
also the variation of $\vrm$ in gray along the lines.}\label{fig2}
\end{figure}

The inflationary potential $\Vhi=\Vhi(z)$ is obtained from
$\vcc(Z)$ in \Eref{vccn} setting $\theta=0$ and $\cm\simeq0$.
I.e.,
\beq \label{vhi} \Vhi=m^2\Om^{-N}\om^{2\nnp}
\lf|U/2\om|^2-3\rg\>\>\mbox{with}\>\>U=\frac{\sqrt{2N}}{J\Om}\Bigg(\lf\sqrt{3}
+\sqrt{N}\rg\Om+2\sqrt{N}\omz\om\Bigg),\eeq
where
$\Om=1-z+16k^2(z-\vrm)^4,~\om=1-z\>\>\mbox{and}\>\>\omz=24k^2(z-v)^3-1/2$.
To localize the position of the inflection point, we impose the
conditions
\beq \Vhi'(z)=\Vhi''(z)=0~~\mbox{for}~~\vrm<z<1,\>\>\mbox{where
$':=d/dz$.}\eeq
For every selected $\vrm$ and $N$ and independently from $m$ these
conditions yield an inflection point $(\ko,\sgo)$. E.g., as shown
in \sFref{fig2}{a}, for $N=12$ and $\vrm=0.5$ we find
$(\ko,\sgo)=(0.40166971,0.707433)$ whereas no inflection point
exists for $k=0.2$ and $k=0.6$. However, varying $N$ and $\vrm$ we
can specify inflection points $(\sgo,\ko)$ for other $k$ too.
E.g., as shown in \sFref{fig2}{b}, for $N=4, 10$ and $30$ (dashed,
solid and dot-dashed line respectively) we show the inflection
points $(\sgo,\ko)$. Along each line we show the variation of
$\vrm$ in gray. Therefore, the presence of inflection point is a
systematic feature of the model.

\subsection{Approaching the Inflationary Dynamics}\label{sec4b}

Due to the complicate form of $\Vhi$ in \Eref{vhi}, we limit
ourselves in expanding numerically $\Vhi$ and $J$ about $z=\sgo$
with results
\beq \Vhi\simeq\vo+\va\dz+\vb\dz^2+\vc\dz^3\>\>\mbox{and}\>\>
J\simeq\jo,\>\>\mbox{where $\dz=z-\sgo$, $\vo=\Vhi(\sgo)$ and
$\jo=J(\sgo)$}\eeq
For the inputs in Fig.~1 the expansion parameters above are given
in \Tref{tab5}. The relevant coefficients depend on the selected
parameters $\vrm, N, k, \sgo$ and $\dk$. Since $\va=\Vhi'(\sgo)$
and $\vb=\Vhi''(\sgo)/2\ll\vo, \vc$ we neglect henceforth terms
with $\va^2, \vb^2$ and $\va\vb$.

\begin{table}[b]\bec \begin{tabular}{|c|c|c|c|c|}\hline
$\vo/(m\mP)^2$&$\va/(m\mP)^2$&$\vb/(m\mP)^2$&$\vc/(m\mP)^2$&$\jo$\\\hline
$3.9\cdot10^{-3}$&$1.5\cdot10^{-6}$&$-2.1\cdot10^{-6}$&$2.2$&$5.4$\\\hline
\end{tabular}\eec
\caption{\sl\small Expansion parameters for the inputs in
Fig.~1.}\label{tab5}
\end{table}

Taking as inputs the parameters above we can investigate the
inflationary evolution by estimating:

\subparagraph{\sf\ftn (a) The slow-roll parameters.} They are
found to be
\beq\epsilon= \left(\frac{\Ve_{{\rm I},\se}}{\sqrt{2}\Ve_{{\rm
I}}}\right)^2\simeq\frac{\va+\dz(2\vb+3\dz\vc)}{\sqrt{2}\jo\vo}~~\mbox{and}~~\eta=\frac{\Ve_{{\rm
I},\se\se}}{\Ve_{\rm I}}\simeq \frac{2(\vb+3\dz\vc)}{\jo^2\vo}\,.
\eeq
The realization of IPI is delimited by the condition
\beq{\ftn\sf max}\{\ep(\se),|\eta(\se)|\}\leq 1,\eeq
which is saturated for $\dz=\dzf$ found as follows
\beq
\eta\lf\dzf\rg\simeq1~~\Rightarrow~~\dzf\simeq-(\jo^2\vo+2\vb)/6\vc<0.
\eeq
Given that $\jo^2\vo\gg\vb$, we expect $\dzf<0$ or $\sgf<\sgx$.

\subparagraph{\sf\ftn (b) The number of e-foldings {\normalsize
$\Ns$} that the scale {\normalsize $\ks=0.05/{\rm Mpc}$}
experiences during IPI.} It is estimated to be
\beq \label{Nse} \Ns=\int_{\sef}^{\sex} d\se\frac{\Vhi}{\Ve_{{\rm
I},\se}}=\frac{\fns-\fnf}{\pn}~~\mbox{where}~~\pn=\frac{\sqrt{3\va\vc}}{\jo^2\vo}\>\>
\mbox{and}\>\>\fn(\sg)=\arctan\frac{\vb+3\sg\vc}{\sqrt{3\va\vc}}.\eeq
Also $\sgx~[\sex]$ is the value of $\sg~[\se]$ when $\ks$ crosses
the inflationary horizon and we define $\fns=\fn(\dzx)$ and
$\fnf=\fn(\dzf)$. Solving it w.r.t $\dzx$ we obtain
\beq \dzx\simeq-\frac{\vb}{3\vc}+\sqrt{\frac{\va}{3\vc}}
\tan\lf\frac{\sqrt{3}\Ns}{\jo^2\vo}+\fnf\rg<0~~\Rightarrow~~\sgx<\sgo.~~\eeq
Therefore, we have no ultra slow roll for $\sgf\leq\sg\leq\sgx$.
Note that $\Ns$, has to be sufficient to resolve the horizon and
flatness problems of standard big bang \cite{plcp} i.e.
\beq \label{Ns} \Ns\simeq61+\ln\lf\pi\vo\Trh^2\rg^{1/6}, \eeq
where $\Trh$ is the reheating temperature -- see below.

\subparagraph{\sf\ftn (c) The amplitude {\normalsize $A_{\rm s}$}
of the power spectrum of the curvature perturbations.} Taking into
account also the normalization of $A_{\rm s}$ \cite{plcp} we
achieve a prediction for the $m$ value. Indeed,
\beq\label{Prob}\As^{1/2}= \frac{1}{2\sqrt{3}\, \pi} \;
\frac{\Vhi^{3/2}(\sex)}{\left|\Ve_{\rm
I,\se}(\sex)\right|}\simeq\frac{\jo\vo^{3/2}}{2\sqrt{3}\pi\va}\cos^{2}\lf\pn\Ns+\fnf\rg\simeq4.588\cdot10^{-5}
~~\Rightarrow~~m\sim10^{-7}\mP.\eeq

\subparagraph{\sf\ftn (d) The remaining inflationary observables.}
Namely, for the {spectral index} $\ns$, its {running}, $\as$, and
the {tensor-to-scalar} ratio $r$ we obtain
\beqs\bea & \hspace*{-.52cm}\ns=\: 1-6\epsilon_\star\ +\
2\eta_\star \simeq 1 +4\pn\tan\lf\pn\Ns+\fnf\rg,\>\>
r=16\epsilon_\star \simeq8\va^2\cos^{-4}\lf\pn\Ns+\fnf\rg/\jo^2\vo^2, ~~~~\label{ns}\\
&\as=\:2\left(4\eta_\star^2-(\ns-1)^2\right)/3-2\xi_\star
\simeq-4\pn\cos^{-2}\lf\pn\Ns+\fnf\rg
\>\>\mbox{with}\>\>\xi={\Ve_{{\rm I},\widehat\sg} \Ve_{{\rm
I},\widehat\sg\widehat\sg\widehat\sg}/\Ve^2_{\rm I}}. \label{as}
\eea\eeqs
Here the variables with subscript $\star$ are evaluated at
$\sg=\sgx$. Note that the combined {\sc Bicep2}/{\slshape Keck
Array} \cite{bk15} and \plk\ data (fitted with the
$\Lambda$CDM$+r+\as$ model) require \cite{plin}
\beq\label{nsex}
\ns=0.9658\pm0.008,\>\as=-0.0066\pm0.014\>\>\mbox{and}\>\>
r\lesssim0.068~~\mbox{at 95\% c.l.}\eeq

For the inputs of Fig. 1 we present some values of the
inflationary parameters in \Tref{tab6} which turn out to be
consistent with \Eref{nsex}. From the values accumulated there, we
observe that the results of our semianalytic approach -- displayed
in curly brackets -- are quite close to the numerical ones. The
semiclassical approximation, used in our analysis, is perfectly
valid since $\Vs^{1/4}\ll\mP$. The $\theta=0$ direction is well
stabilized and does not contribute to the curvature perturbation,
since for the relevant effective mass $m_{\theta\rm I}$ we find
$m^2_{\theta\rm I}>0$ for $N>3$ and $m_{\theta\rm I\star}/\Hs>1$
where $\Hhi=(\Vhi/3)^{1/2}$. The one-loop radiative corrections,
$\dV$, to $\Vhi$ induced by $m_{\theta\rm I}$ let intact our
inflationary outputs.

\begin{table}[t]\renewcommand{\arraystretch}{1.2}
\bec \begin{tabular}{|c|c|c|c|c|}\hline
$\ko/0.1$&$\sgo/0.1\mP$&$\dk/10^{-6}$&\multicolumn{2}{c|}{$\dzx/10^{-4}\mP$}\\\hline
$4.0166971$&$7.07433$&$3.20232$&\multicolumn{2}{c|}{$-1.5$~\{$-1.1$\}}\\\hline
$\Vs^{1/4}/\EeV$&$\Hs/\EeV$&\multicolumn{2}{c|}{$\dzf/10^{-2}\mP$}&$m_{\theta\rm
I\star}/\Hs$\\\hline
$4.6\cdot10^{5}$&$49.5$&\multicolumn{2}{c|}{$-1.16$~\{$-0.87$\}}&$5.1$\\\hline
$\ns$&$r/10^{-8}$&$-\as/10^{-3}$&$10^5\As^{1/2}$&$\Ns$\\\hline
$0.966$~\{$0.97$\}&$4.8$~\{$3.9$\}&$3.3$~\{$3.2$\}&$4.59$~\{$4.27$\}&$46.5$~\{$45$\}\\\hline
\end{tabular}\eec\renewcommand{\arraystretch}{1.}
\caption{\sl\small Sample values of inflationary parameters for
the inputs in Fig.~1.}\label{tab6}
\end{table}

\subsection{Inflaton Decay and Reheating}\label{sec4c}

Soon after the end of IPI, the (canonically normalized) sgoldstino
\beq\dzvh=\vev{J}\dzv~~\mbox{with}~~\dzv=z-\vrm~~\mbox{and}~~
\vev{J}=\sqrt{\frac{N}{2}}\frac{1}{\vev{\om}} \label{dzv} \eeq
settles into a phase of damped oscillations abound the minimum
reheating the universe at a temperature
\beq\label{Trh} \Trh= \left({72/5\pi^2g_{\rm
rh*}}\right)^{1/4}\Gsn^{1/2}\mP^{1/2}\>\>\mbox{where $g_{\rm
rh*}=106.75$ and}\>\>\>\Gsn\simeq\Ggr+\Gth+\Gh\eeq
the total decay width, $\Gsn$, of $\dzvh$ with the individual
decay widths are found to be
\beq \Ggr\simeq\frac{\vev{\om}^{-\sqrt{3N}}\msn^5}{96\pi
m^2\mP^2},~~\Gth\simeq\frac{\msn^3}{16N\pi
\vrm\mP}~~\mbox{and}~~\Gh=\frac{N\what\mu^2}{16\pi\mP^2}\msn\,.\label{Gs}\eeq
They express decay of $\dzvh$ into gravitinos, pseudo-sgoldstinos
and higgsinos via the $\mu$ term respectively. Note that $\Gh$
becomes rather enhanced for large $N$'s. Thanks to the high $\msn$
and $\what\mu$ values, no moduli problem arises in this context
since $\Trh\sim1~\PeV\gg1~{\rm MeV}$.

\section{Results}\label{res}

The free parameters of the model are
$$m, N, \vrm, \dk=k-\ko~~\mbox{and}~~\dzx=\sgx-\sgo$$
Recall that $(\ko,\sgo)$ is the inflection point which can be
computed self-consistently for any selected $N$ and $\vrm$.
Enforcing \eqs{Ns}{Prob} we restrict $\dk$ and $m$ whereas the
$\ns$ bounds in \Eref{nsex} determines $\dzx$. Increasing $\dk$
allows us to increase the slope of the plateau around $\sgo$
decreasing, thereby, $\Ns$. The model's predictions regard $\as$
and $r$ estimated from \Eref{as}.

\begin{figure}[t]\vspace*{-.12in}
\hspace*{-.12in}
\begin{minipage}{8in}
\epsfig{file=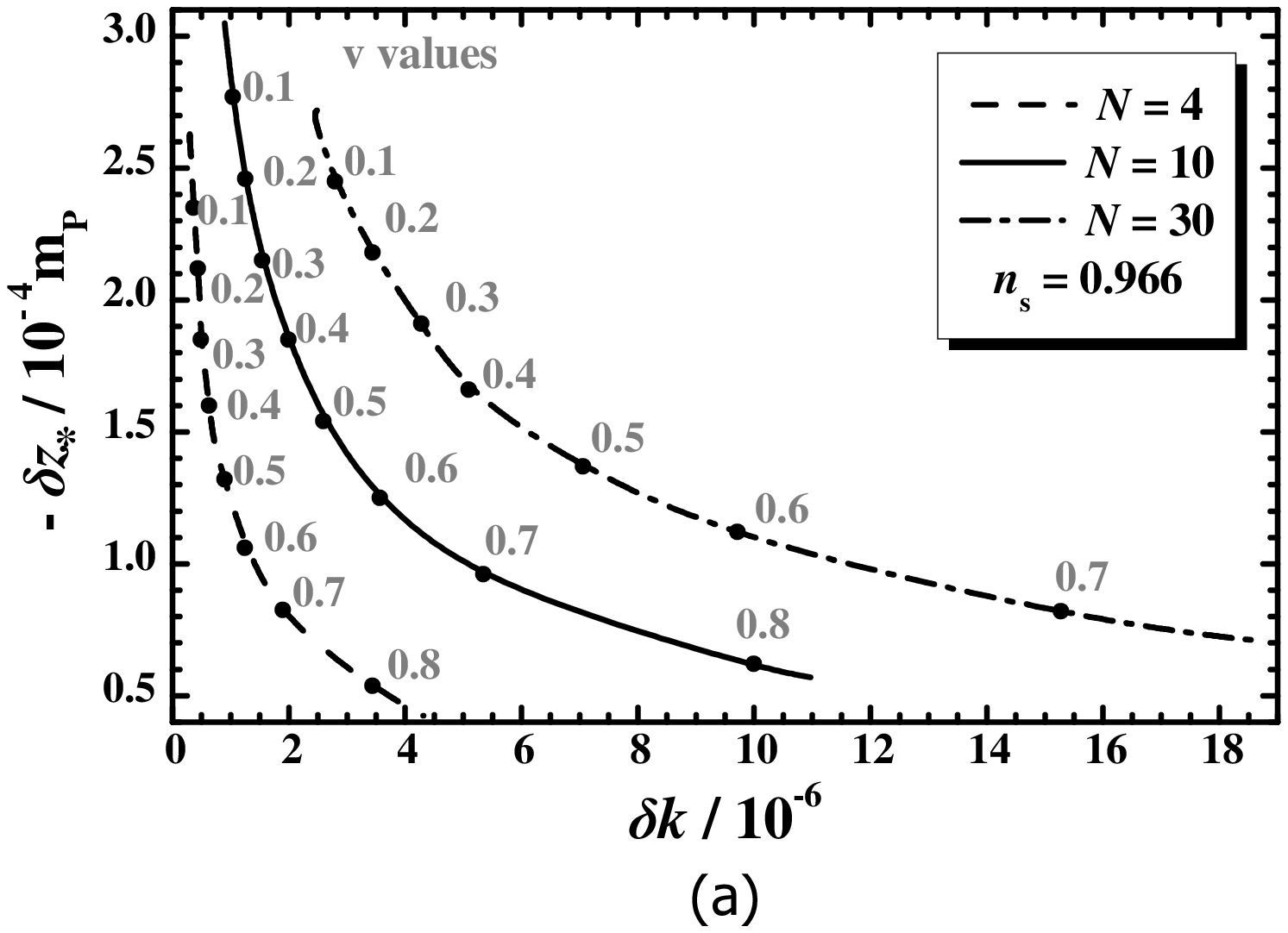,height=3.3in,angle=-90} \hspace*{-.2cm}
\epsfig{file=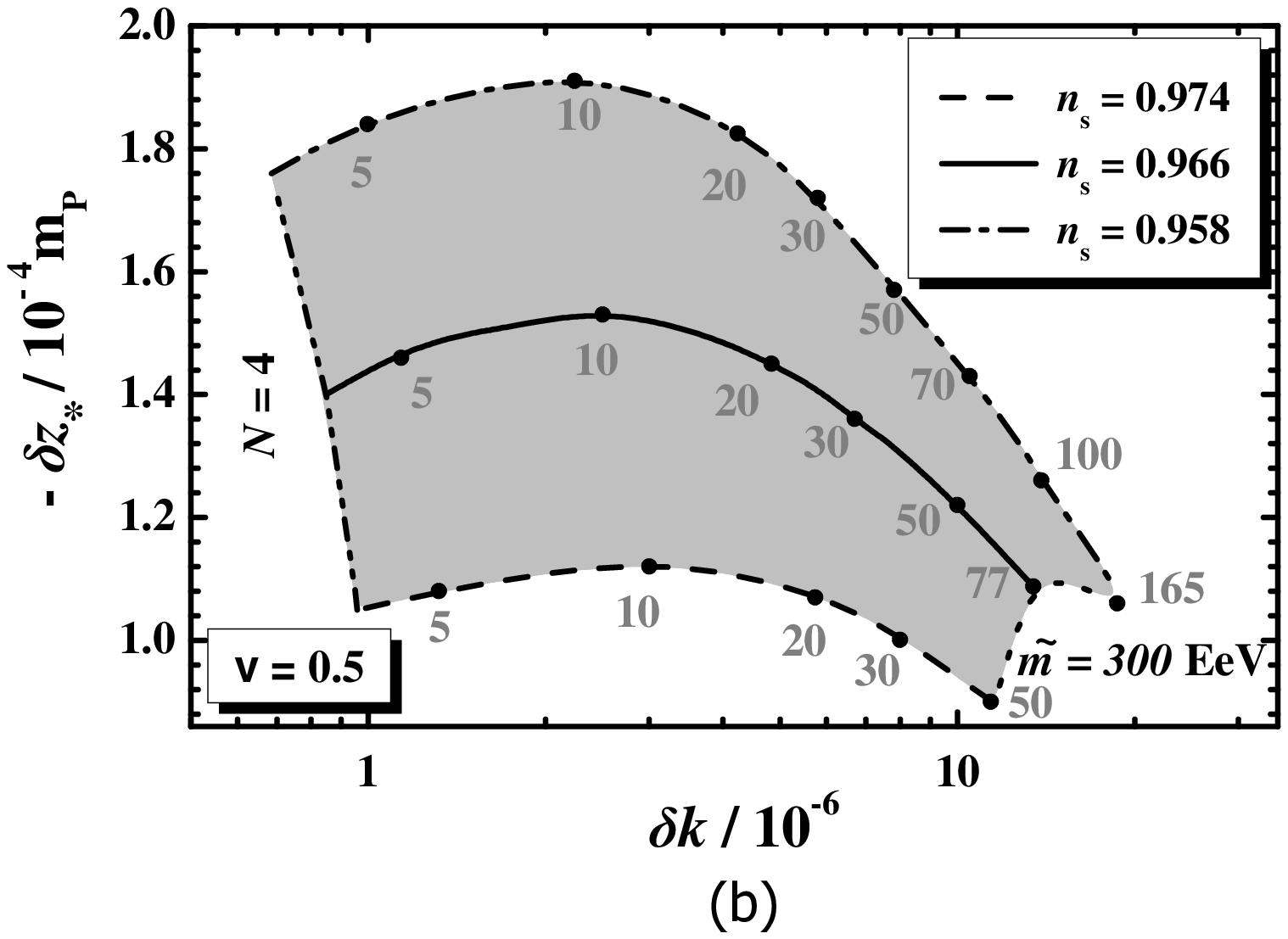,height=3.3in,angle=-90} \hfill
\end{minipage}
\hfill \caption{\sl\small In the $\dk-(-\dzx)$ plane we present
({\sf\ftn a}) allowed curves  for $\ns\simeq0.966$ and various
$N$'s and ({\sf\ftn b}) allowed (shaded) region varying
continuously $N$. The variation of ({\sf\ftn a}) $\vrm$ or
({\sf\ftn b}) $N$ is shown in gray along the various lines. The
constraint fulfilled along each line is shown in
black.}\label{fig3}
\end{figure}

The outputs of our numerical investigation are presented:

\subparagraph{\sf\ftn (a) In Fig.~3, where we plot allowed domains
in the {\normalsize $\dk-(-\dzx)$} plane.} In \sFref{fig3}{a} we
fix $N$ to three representative values $4, 10$ and $30$ and
display the allowed curves (dot-dashed, solid and dashed lines
respectively) taking the central $\ns$ value in \Eref{nsex}. The
variation of $\vrm$ along each line is displayed in gray.  On the
other hand, in \sFref{fig3}{b} we set $\vrm=0.5$ and identify the
allowed (shaded) region by varying $\ns$ in the margin of
\Eref{nsex}. The variation of $N$ is shown along each line.
Besides the bounds on $\ns$ in \Eref{nsex}, which yield the dashed
and dot-dashed lines, we take into account the upper bound in
\Eref{highb} which is saturated along the dotted line and the
lower bound on $N$, mentioned above \Eref{msa}, along the double
dotted dashed line. We remark that increasing $|\dzx|$, decreases
$\ns$ with fixed $\dk$. For $N=10$ and central $\ns$ in
\Eref{nsex} the inflationary predictions are
\beq\as\simeq-3\cdot10^{-3}\>\>\>\mbox{and}\>\>\>r\simeq5\cdot10^{-8}\>\>\>
\mbox{for}\>\>\>\Trh\simeq1.75~\PeV\>\>\>\mbox{and}\>\>\>\Ns\simeq46.5\eeq
The obtained $\as$ might be detectable in the future
\cite{asdrees}. The needed tuning is of the order of $10^{-6}$
which is certainly ugly but milder than that needed for IPI within
the conventional MSSM \cite{ipisusy}.

\begin{figure}[!t]\vspace*{-.12in}
\hspace*{-.12in}
\begin{minipage}{8in}
\epsfig{file=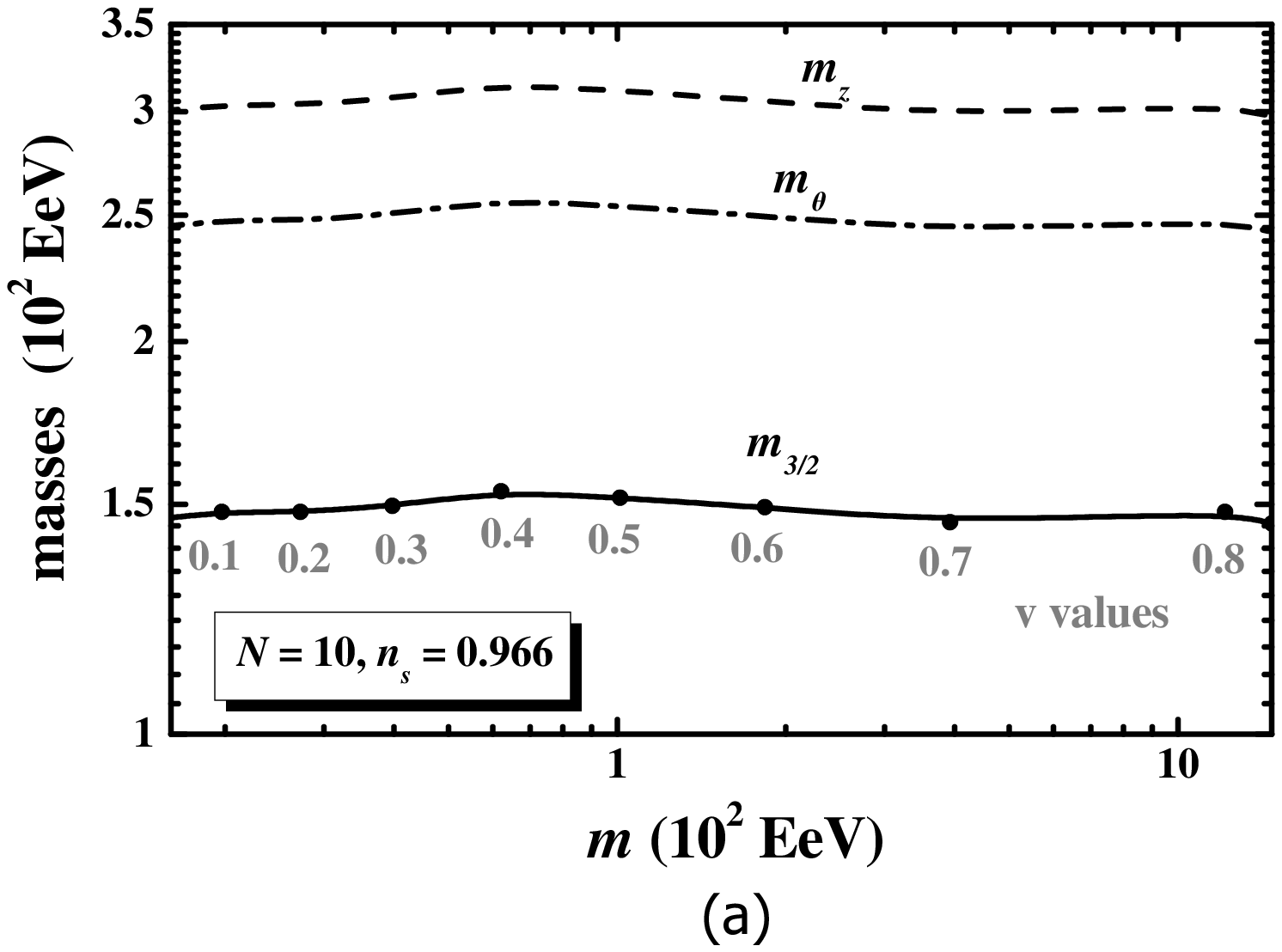,height=3.3in,angle=-90}
\hspace*{-.2cm}
\epsfig{file=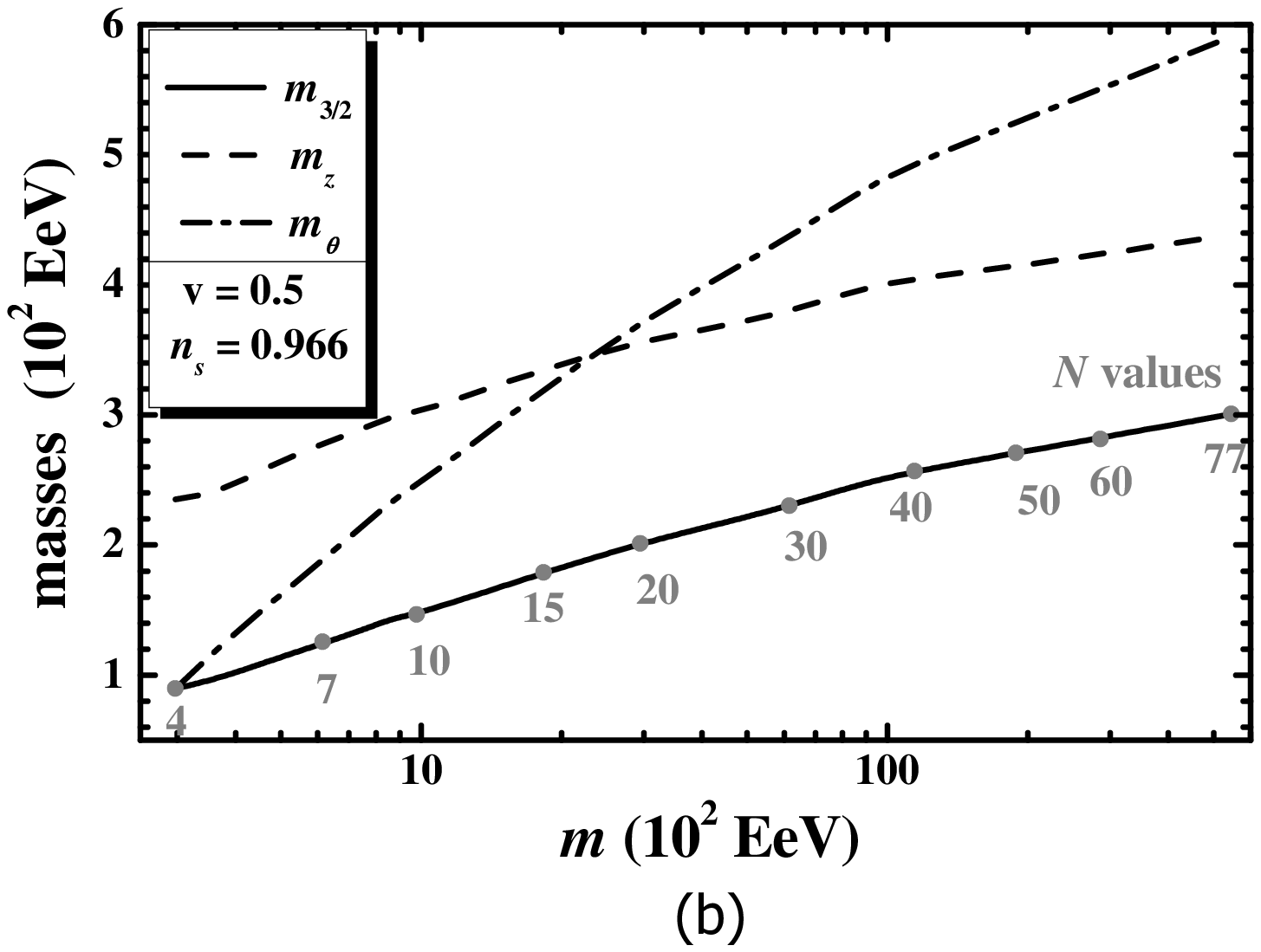,height=3.3in,angle=-90} \hfill
\end{minipage}
\hfill \caption{\sl\small Allowed values of $\mgr$, $\mz$ and
$\mth$ (solid, dashed and dot-dashed line respectively) versus $m$
for $\ns\simeq0.966$ and ({\sf\ftn a}) $N=10$ or ({\sf\ftn b})
$\vrm=0.5$. The variation of $\vrm$ ({\sf\ftn a}) or ({\sf\ftn b})
$N$ is shown along the solid line in gray.}\label{fig4}
\end{figure}

\subparagraph{\sf\ftn (b) In Fig.~4, where we plot the mass
spectrum as a function of {\normalsize $m$} for {\normalsize
$\ns\simeq0.966$}.} Namely, in \sFref{fig4}{a} we fix $N=10$ and
vary $\vrm$ whereas in \sFref{fig4}{b} we fix $\vrm=0.5$ and vary
$N$. The variation of the variable parameter is shown along the
solid line in gray. The allowed values of $\mgr$, $\mz$ and
$\mth$, estimated by the expression in \Eref{msa}, are depicted by
a solid, dashed and dot-dashed line respectively. From
\sFref{fig4}{a} we remark that the hierarchy of the particle
masses remains constant for fixed $N$. They remain of the order of
$100~\EeV$ whereas $m$ becomes larger and larger than this level
as $\vrm$ and/or $N$ increases. This is explained from \Eref{msa}
if we take into account that $\om<1$ from \Eref{om} and
$\sqrt{3N}/2>1$. In total we obtain
\beq 3\lesssim \frac{m}{1~\EeV}\lesssim55600,~~0.89\lesssim
\frac{\mgr}{100~\EeV}\lesssim3,\>\> 2.3\lesssim
\frac{\msn}{100~\EeV}\lesssim4.4,~~\mbox{and}~~0.89\lesssim
\frac{\mth}{100~\EeV}\lesssim59.\eeq
This mass spectrum hints towards high-scale MSSM consistently with
the LHC results on the Higgs boson mass -- see \Eref{highb}.
Needless to say, the stability of the electroweak vacuum up to
$\mP$ is automatically assured within this framework
\cite{strumia}. On the other hand, the gauge hierarchy problem
becomes acute since the SUSY-mass scale is much higher than the
electroweak scale and the relevant fine-tuning needed remains
unexplained. From \sFref{fig4}{b} we also infer that the $\dzvh$
decay channel into $\theta$'s is kinematically blocked for
$N\gtrsim20$. We also find that for $N\lesssim10$,
$\Gsn\simeq\Ggr+\Gh$ whereas for larger $N$'s $\Gsn\simeq\Gh$ and
so \Eref{Trh} yields $\Trh\simeq(4-20)~\PeV$ resulting to
$\Ns\simeq(45.5-46.7)$.

\newpage
\section{Conclusions}\label{con}

We proposed a SUGRA model with just one gauge-singlet chiral
superfield (the Goldstino) that offers at once:

\begin{itemize}

\item Tiny cosmological constant in the low-energy vacuum at the
cost of a fine tuned parameter;

\item Inflection-point inflation resulting to an adjustable $\ns$,
a small $r$ and a sizable $\as\sim-10^{-3}$;

\item Spontaneous SUSY breaking at the scale $\mss\sim 100~\EeV$,
which is consistent with the Higgs boson mass measured at LHC
within high-scale SUSY.

\end{itemize}

It would be interesting to investigate the following issues:

\begin{itemize}

\item The generation of primordial black holes, which is currently
under debate \cite{yoko,riotto} during an ultra slow-roll phase.
Here we did not address the question of how $\sg$ reaches $\sgo$.
Since $\sgx<\sgo$, we assumed that the slow-roll approximation
offers a reliable description of IPI. This is true if $\sg$ lies
initially near $\sgo$ with a small enough kinetic energy density
\cite{dim}.

\item The candidacy of intermediate-scale lightest neutralino with
mass $M_1\sim\EeV$ in the interval $\Trh<M_1<T_{\rm max}$ as a
cold dark matter candidate adapting the production mechanism of
WIMPZILLAS \cite{kolb}.

\item The realization baryogenesis via non-thermal leptogenesis
taking into account similar attempts -- see e.g. \cref{ntlepto}.

\item The reconciliation of our proposal with swampland
conjectures \cite{vafa} -- see e.g. \cref{sevilla} for relevant
modifications which may render our setting more friendly with the
string ultraviolet completions.

\end{itemize}

\paragraph*{\small \bf\scshape Acknowledgments} {\small I would like to thank S.
Ketov and M. Saridakis and for interesting discussions. This
research work was supported by the Hellenic Foundation for
Research and Innovation (H.F.R.I.) under the ``First Call for
H.F.R.I. Research Projects to support Faculty members and
Researchers and the procurement of high-cost research equipment
grant'' (Project Number: 2251).}

\def\ijmp#1#2#3{{\emph{Int. Jour. Mod. Phys.}}
{\bf #1},~#3~(#2)}
\def\plb#1#2#3{{\emph{Phys. Lett.  B }}{\bf #1},~#3~(#2)}
\def\zpc#1#2#3{{Z. Phys. C }{\bf #1},~#3~(#2)}
\def\prl#1#2#3{{\emph{Phys. Rev. Lett.} }
{\bf #1},~#3~(#2)}
\def\rmp#1#2#3{{Rev. Mod. Phys.}
{\bf #1},~#3~(#2)}
\def\prep#1#2#3{\emph{Phys. Rep. }{\bf #1},~#3~(#2)}
\def\prd#1#2#3{{\emph{Phys. Rev.  D }}{\bf #1},~#3~(#2)}
\def\npb#1#2#3{{\emph{Nucl. Phys.} }{\bf B#1},~#3~(#2)}
\def\npps#1#2#3{{Nucl. Phys. B (Proc. Sup.)}
{\bf #1},~#3~(#2)}
\def\mpl#1#2#3{{Mod. Phys. Lett.}
{\bf #1},~#3~(#2)}
\def\arnps#1#2#3{{Annu. Rev. Nucl. Part. Sci.}
{\bf #1},~#3~(#2)}
\def\sjnp#1#2#3{{Sov. J. Nucl. Phys.}
{\bf #1},~#3~(#2)}
\def\jetp#1#2#3{{JETP Lett. }{\bf #1},~#3~(#2)}
\def\app#1#2#3{{Acta Phys. Polon.}
{\bf #1},~#3~(#2)}
\def\rnc#1#2#3{{Riv. Nuovo Cim.}
{\bf #1},~#3~(#2)}
\def\ap#1#2#3{{Ann. Phys. }{\bf #1},~#3~(#2)}
\def\ptp#1#2#3{{Prog. Theor. Phys.}
{\bf #1},~#3~(#2)}
\def\apjl#1#2#3{{Astrophys. J. Lett.}
{\bf #1},~#3~(#2)}
\def\n#1#2#3{{Nature }{\bf #1},~#3~(#2)}
\def\apj#1#2#3{{Astrophys. J.}
{\bf #1},~#3~(#2)}
\def\anj#1#2#3{{Astron. J. }{\bf #1},~#3~(#2)}
\def\mnras#1#2#3{{MNRAS }{\bf #1},~#3~(#2)}
\def\grg#1#2#3{{Gen. Rel. Grav.}
{\bf #1},~#3~(#2)}
\def\s#1#2#3{{Science }{\bf #1},~#3~(#2)}
\def\baas#1#2#3{{Bull. Am. Astron. Soc.}
{\bf #1},~#3~(#2)}
\def\ibid#1#2#3{{\it ibid. }{\bf #1},~#3~(#2)}
\def\cpc#1#2#3{{Comput. Phys. Commun.}
{\bf #1},~#3~(#2)}
\def\astp#1#2#3{{Astropart. Phys.}
{\bf #1},~#3~(#2)}
\def\epjc#1#2#3{{Eur. Phys. J. C}
{\bf #1},~#3~(#2)}
\def\nima#1#2#3{{Nucl. Instrum. Meth. A}
{\bf #1},~#3~(#2)}
\def\jhep#1#2#3{{\emph{J. High Energy Phys.} }
{\bf #1},~#3~(#2)}
\def\jcap#1#2#3{{\emph{J. Cosmol. Astropart. Phys.} }
{\bf #1},~#3~(#2)}
\def\jcapn#1#2#3#4{{\sl J. Cosmol. Astropart. Phys. }{\bf #1}, no. #4, #3 (#2)}
\def\prdn#1#2#3#4{{\sl Phys. Rev. D }{\bf #1}, no. #4, #3 (#2)}
\newcommand{\arxiv}[1]{{\ftn\tt  arXiv:#1}}
\newcommand{\hepph}[1]{{\ftn\tt  hep-ph/#1}}
\newcommand{\hepth}[1]{{\ftn\tt  hep-th/#1}}
\newcommand{\astroph}[1]{{\ftn\tt  astro-ph/#1}}
\def\prdn#1#2#3#4{{\sl Phys. Rev. D }{\bf #1}, no. #4, #3 (#2)}
\def\jcapn#1#2#3#4{{\sl J. Cosmol. Astropart.
Phys. }{\bf #1}, no. #4, #3 (#2)}
\def\epjcn#1#2#3#4{{\sl Eur. Phys. J. C }{\bf #1}, no. #4, #3 (#2)}


\begin{thebibliography}{99}



\bibitem{ns89}  J.~Ellis, B.~Nagaraj, D.V.~Nanopoulos and
K.A.~Olive, \emph{De Sitter Vacua in No-Scale Supergravity},
\jhep{11}{2018}{110} [\arxiv{1809.10114}]; J.~Ellis, B.~Nagaraj,
D.V.~Nanopoulos, K.A.~Olive and S.~Verner, \emph{From Minkowski to
de Sitter in Multifield No-Scale Models}, \jhep{10}{2019}{161}
[\arxiv{1907.09123}].


\bibitem{de} C.~Pallis, {\it From Minkowski to de Sitter Vacua with Various Geometries},
{\sl Eur. Phys. J. C }\textbf{83}, no.~4, 328 (2023)
[\arxiv{2211.05067}].


\bibitem{dei} C.~Pallis, {\it Inflection-point sgoldstino inflation in no-scale
supergravity,} {\sl Phys. Lett. B} {\bf 843}, 138018 (2023)
[\arxiv{2302.12214}].




\bibitem{plcp} N.~Aghanim {\it et al.} [\plk\ Collaboration],
{\it Planck 2018 results. VI. Cosmological parameters}, {\sl
Astron. Astrophys. }\textbf{641}, A6 (2020) [\arxiv{1807.06209}].
%
\bibitem{plin} Y.~Akrami {\it et al.} [\plk\ Collaboration], {\it Planck 2018 results.
X. Constraints on inflation}, {\sl Astron. Astrophys.
}\textbf{641}, A10 (2020) [\arxiv{1807.06211}].

\bibitem{bk15} P.A.R. Ade \etal, {\it BICEP2 / Keck Array x: Constraints
on Primordial Gravitational Waves using Planck, WMAP, and New
BICEP2/Keck Observations through the 2015 Season}, {\sl Phys. Rev.
Lett.} {\bf 121}, 221301 (2018) [\arxiv{1810.05216}].


\bibitem{polonyi} J. Polonyi, {\it Generalization of the massive scalar
multiplet coupling to the supergravity}, Budapest preprint
KFKI/1977/93 (1977).

\bibitem{hall} M. Claudson, L. Hall and I. Hinchliffe,
{\it Tuning the cosmological constant in N=1 supergravity with an
$R$ symmetry}, {\sl Phys. Lett. B }{\bf 130}, 260 (1983).

\bibitem{kallosh} S. Kachru, R. Kallosh, A.D. Linde and S. P. Trivedi, {\it De
Sitter vacua in string theory}, {\sl Phys. Rev. D} {\bf 68},
046005 (2003) [\hepth{0301240}].



\bibitem{susyr} C.~Pallis, {\it Gravity-mediated SUSY breaking, R symmetry,
and hyperbolic K\"ahler geometry,} {\sl Phys.\ Rev.\ D }{\bf 100},
no.~5, 055013 (2019) [\arxiv{1812.10284}]; C.~Pallis, {\it
SUSY-breaking scenarios with a mildly violated $R$ symmetry,} {\sl
Eur. Phys. J. C} \textbf{81}, no.~9, 804 (2021)
[\arxiv{2007.06012}].

\bibitem{nsreview} J.~Ellis, M.A.G.~Garcia, N.~Nagata, D.V. Nanopoulos, K.A.~Olive
and S.~Verner, {\it Building models of inflation in no-scale
supergravity}, {\sl Int. J. Mod. Phys. D} {\bf 29},  16, 2030011
(2020) [\arxiv{2009.01709}].




\bibitem{symm} G.~'t Hooft, \emph{Naturalness, chiral symmetry,
and spontaneous chiral symmetry breaking}, {\sl NATO Sci.\ Ser.\ B
}{\bf 59}, 135 (1980).








\bibitem{ant1} R. Kallosh and A. Linde, {\it Planck, LHC and
$\alpha$-attractors,} {\sl Phys. Rev. D} {\bf 91}, 083528 (2015)
[\arxiv{1502.07733}]; M.C.~Rom\~ao and S.F.~King, {\it
Starobinsky-like inflation in no-scale supergravity Wess-Zumino
model with Polonyi term} \jhep{07}{2017}{033}
[\arxiv{1703.08333}]; K. Harigaya and K. Schmitz, {\it Inflation
from High-Scale Supersymmetry Breaking,} {\sl Phys. Lett. B} {\bf
773}, 320 (2017) [\arxiv{1707.03646}]; I.~Antoniadis,
A.~Chatrabhuti, H.~Isono and R.~Knoops, {\it Inflation from
Supersymmetry Breaking,} {\sl Eur. Phys. J. C }\textbf{77},
no.~11, 724 (2017) [\arxiv{1706.04133}]; E.~Dudas, T.~Gherghetta,
Y.~Mambrini and K.A.~Olive, {\it Inflation and High-Scale
Supersymmetry with an EeV Gravitino}, {\sl Phys.\ Rev.\ D } {\bf
96}, no. 11, 115032 (2017) [\arxiv{1710.07341}]; Y.~Aldabergenov,
A.~Chatrabhuti and H.~Isono, {\it $\alpha$-attractors from
supersymmetry breaking,} {\sl Eur. Phys. J. C }\textbf{81}, no.~2,
166 (2021) [\arxiv{2009.02203}].




%


\bibitem{ipisusy} R. Allahverdi, K. Enqvist, J. Garcia-Bellido and A. Mazumdar,
\emph{Inflection point inflation within supersymmetry}, {\sl Phys.
Rev. Lett. }{\bf 97}, 191304 (2006) [\hepph{0605035}]; J.C. Bueno
Sanchez, K. Dimopoulos and D.H. Lyth, {\it A-term inflation and
the MSSM}, \jcap{01}{2007}{015} [\hepph{0608299}].


\bibitem{drees} S.-M. Choi and H.M. Lee, {\it Inflection point inflation and reheating}, {\sl Eur. Phys. J. C} {\bf 76} 303, no. 6 (2016)
[\arxiv{1601.05979}]; M.~Drees and Y.~Xu, {\it Small field
polynomial inflation: reheating, radiative stability and lower
bound}, \jcap{09}{2021}{012} [\arxiv{2104.03977}].

\bibitem{epole} C.~Pallis, \emph{An alternative framework for E-model inflation in supergravity},
{\sl Eur. Phys. J. C}~{\bf 82}, no.~5, 444 (2022)
[\arxiv{2204.01047}].



\bibitem{asdrees} J.B.~Mu\~noz \etal, \emph{Towards a measurement of the spectral
runnings}, \jcap{05}{2017}{032} [\arxiv{1611.05883}].


\bibitem{strumia} E.~Bagnaschi, G.F.~Giudice,
P.~Slavich and A.~Strumia, {\it Higgs Mass and Unnatural
Supersymmetry}, \jhep{09}{2014}{092} [\arxiv{1407.4081}].


\bibitem{soft} A. Brignole, L.E. Ib\'a\~nez and C. Mu\~noz, {\it Soft supersymmetry
breaking terms from supergravity and superstring models}, {\sl
Adv.\ Ser.\ Direct.\ High Energy Phys. }{\bf 18}, 125 (1998)
[\hepph{9707209}].


\bibitem{rh} C. Pallis, {\it Kination-dominated reheating and cold dark matter abundance}, \npb{751}{2006}{129}
[\hepph{0510234}]; J.~Ellis \etal, {\it BICEP/Keck constraints on
attractor models of inflation and reheating}, {\sl Phys. Rev. D}
\textbf{105}, no.4, 043504 (2022) [\arxiv{2112.04466}].

\bibitem{rhdecay} M.~Endo, F.~Takahashi and T.T.~Yanagida, \emph{Inflaton Decay in Supergravity},
{\sl Phys. Rev. D }\textbf{76}, 083509 (2007) [\arxiv{0706.0986}];
J. Ellis, M. Garcia, D. Nanopoulos and K. Olive,
\emph{Phenomenological Aspects of No-Scale Inflation Models}, {\sl
J. Cosmol. Astropart. Phys. }{\bf 10}, 003 (2015)
[\arxiv{1503.08867}]; Y.~Aldabergenov, I.~Antoniadis,
A.~Chatrabhuti and H.~Isono, \emph{Reheating after inflation by
supersymmetry breaking}, {\sl Eur. Phys. J. C }\textbf{81},
no.~12, 1078 (2021) [\arxiv{2110.01347}]; K.J.~Bae, H.~Baer,
V.~Barger and R.W.~Deal, \emph{The cosmological moduli problem and
naturalness}, \jhep{02}{2022}{138} [\arxiv{2201.06633}].





\bibitem{bh} J. Garcia-Bellido and E. Ruiz Morales, {\it Primordial black holes from single field
models of inflation}, {\sl Phys. Dark Univ. }{\bf 18}, 47 (2017)
[\arxiv{1702.03901}]; C. Germani and T. Prokopec, {\it On
primordial black holes from an inflection point}, {\sl Phys. Dark
Univ. }{\bf 18}, 6 (2017) [\arxiv{1706.04226}].


\bibitem{dim}  K. Dimopoulos, {\it Ultra slow-roll inflation demystified},
{\sl Phys. Lett. B}{\bf 775}, 262 (2017)[\arxiv{1707.05644}].


\bibitem{yoko} J.~Kristiano and J.~Yokoyama,
{\it Ruling Out Primordial Black Hole Formation From Single-Field
Inflation,} \arxiv{2211.03395}.

\bibitem{riotto} H.~Firouzjahi and A.~Riotto, {\it Primordial Black Holes and loops
in single-field inflation,} \jcap{02}{2024}{021}
[\arxiv{2304.07801}].




\bibitem{kolb} D.J.H.~Chung, E.W.~Kolb and A.~Riotto,
\emph{Nonthermal supermassive dark matter}, {\sl Phys. Rev. Lett.
}\textbf{81}, 4048 (1998) [\hepph{9805473}].


\bibitem{ntlepto} K.~Kaneta, Y.~Mambrini, K.A.~Olive and S.~Verner,
{\it Inflation and Leptogenesis in High-Scale Supersymmetry}, {\sl
Phys. Rev. D }\textbf{101}, no.~1, 015002 (2020)
[\arxiv{1911.02463}].



\bibitem{vafa} C. Vafa,
{\it Distance and de Sitter conjectures on the swampland},
\hepth{0509212}.

%


\bibitem{sevilla} I.M.~Rasulian, M.~Torabian and L.~Velasco-Sevilla, {\it Swampland de Sitter conjectures
in no-scale supergravity models}, {\sl Phys. Rev. D }\textbf{104},
no.4, 044028 (2021) [\arxiv{2105.14501}].




\end{thebibliography}
\end{document}